\documentclass[a4paper,11pt]{article}
\pdfoutput=1 

\usepackage{jcappub} 

\usepackage[T1]{fontenc} 

\title{\boldmath Installing of cosmological constant}

\author[a,b]{V.V. Kiselev}

\affiliation[a]{Russian State Research Center
Institute for High Energy Physics (National Research Centre Kurchatov
Institute), Russia, 142281, Moscow Region, Protvino, Nauki 1}
\affiliation[b]{Moscow Institute of Physics and Technology (State University),
Russia, 141701, Moscow Region, Dolgoprudny, Institutsky 9}

\emailAdd{Valery.Kiselev@ihep.ru}

\abstract{An artificial scale of observable cosmological constant is
dynamically related to its natural bare value due to an installation of
relevant coherent state for the inflationary field in a finite volume of
early Universe, because of exponential suppression of probability to find the
state with zero number of quanta. Homogeneous quantum fluctuations of the
field actually put hard constraints on the total amount of inflation.}

\begin{document}
\maketitle
\flushbottom

\section{Introduction}
\label{sec:intro}

Since the Universe inflation
\cite{Star-1,MCh,i-Guth,i-Linde,i-Albrecht+Steinhardt,i-Linde2,inflation}
started with a final volume $V_R$, a homogeneous inflationary field $\phi(t)$
had got nonzero quantum fluctuations of both its value $\delta \phi(t)$ and
rate $\delta\dot\phi(t)$. Then the inflaton energy-momentum tensor
$T_\mu^\nu$ was averaged over a quantum state
$$
    |\alpha\rangle=|0\rangle\langle 0|\alpha\rangle+|q\rangle\langle
    q|\alpha\rangle,
$$
where $|0\rangle$ is the vacuum with the fluctuations relevant to the volume
$V_R$, while $|q\rangle$ represents the sum over states with nonzero number
of $\phi$ quanta, so that the probability  of vacuum $w_\mathrm{vac}=|\langle
0|\alpha\rangle|^2$ in this state of inflation installation
$$
    w_\mathrm{vac}\sim\mathrm{e}^{-\ell},
$$
with $\ell$ being the average number of quanta in the final volume
$V_R<\infty$, hence, $\ell<\infty$ and
\begin{equation}\label{eq:T}
    \langle \alpha|T_\mu^\nu|\alpha\rangle=
    w_\mathrm{vac}\langle 0|T_\mu^\nu|0\rangle+
    (1-w_\mathrm{vac})\langle q|T_\mu^\nu|q\rangle+
    \sqrt{w_\mathrm{vac}(1-w_\mathrm{vac})}\left\{
    \langle 0|T_\mu^\nu|q\rangle+\langle q|T_\mu^\nu|0\rangle
    \right\}.
\end{equation}
Here
$$
    \langle 0|T_\mu^\nu|0\rangle=\rho_\Lambda^\mathrm{bare}\delta_\mu^\nu
$$
is the bare cosmological term with the vacuum energy
$\rho_\Lambda^\mathrm{bare}$ \cite{Peebles,SW}, for instance,
\begin{equation}\label{eq:bare}
    \rho_\Lambda^\mathrm{bare}=\frac12\langle 0|\dot\phi^2|0\rangle+
    \frac{m^2}{2}\langle 0|\phi^2|0\rangle=\frac12\left(
    \delta\dot\phi^2_\mathrm{v}+m^2\delta\phi^2_\mathrm{v}\right)
\end{equation}
in the simplest model of inflaton with a mass $m$. Thus, the observed
cosmological constant gets an artificial value
$$
    \rho_\Lambda\sim \mathrm{e}^{-\ell} \rho_\Lambda^\mathrm{bare}\ll
    \rho_\Lambda^\mathrm{bare}\qquad\mbox{at }\ell\gg1.
$$
In (\ref{eq:T}) the second term represents the dynamical contribution of
inflaton quanta, while the third term corresponds to the cosmological
creation (and annihilation) of those quanta from (and to) the vacuum due to
the source of gravitational interaction, i.e. due to the energy-momentum
tensor.

The bare cosmological constant can be evaluated in the following way:

It is natural to set the quantum fluctuations of kinetic and potential
energies to be comparable for the inflationary field in the final volume
$V_R$ at the inflation start, hence, $\delta\dot\phi\sim m\,\delta\phi.$ For
inflaton potential $V(\phi)$ the slow roll regime of field evolution at
Hubble rate $H$ gives
$$
    \dot\phi\sim\frac{\partial V}{\partial \phi}\,\frac{1}{H}
    \sim\tilde m^2\phi\,\frac{\tilde m_\mathrm{Pl}}{m\phi}\sim m\,
    \tilde m_\mathrm{Pl},
$$
where $\tilde m_\mathrm{Pl}^2=(8\pi G)^{-1}$ is the reduced Planck mass
squared. The slow rolling can be installed only if the rate $\dot\phi$ begins
to exceed its fluctuation $\delta\dot\phi$, i.e. at $\dot\phi\sim\delta\dot
\phi$. Therefore,
$$
    \rho_\Lambda^\mathrm{bare}\sim\delta\dot\phi^2\sim\dot\phi^2\sim
    m^2\tilde m_\mathrm{Pl}^2.
$$
Empirically, $m\sim 10^{13}$ GeV \cite{Bstrap}, so the value of
$\rho_\Lambda^\mathrm{bare} \sim(10^{15}\mbox{ Gev})^4$ that is quite natural
for high energy particle physics.

In this paper we consider the role of quantum fluctuations for the
homogeneous inflationary field with the potential
\begin{equation}\label{m2}
    V=\frac12\, m^2\phi^2,
\end{equation}
in order to derive the condition of $\delta\dot\phi^2=\dot\phi^2$ for the
inflation start. We also argue for the coherent state of inflaton at the
moment of inflation installation. For the number of inflationary field quanta
in the primary volume $V_R$ of inflation start the estimate $\ell\sim 250$ is
obtained.

\section{Problem treatment}

As it was discovered by measuring magnitudes of brightness for standard
candles of type Ia supernovas versus its red shifts
 \cite{Riess,SN1,SN2,SN3,SN4}, the accelerated expansion of Universe can be
explained by the cosmological constant (see  \cite{Peebles}), which was
invented by Albert Einstein for effects of anti-gravity. So, the cosmological
term determines the vacuum energy density that is given by fourth degree of
scale about $10^{-3}$ eV, corresponding to 1 eV per cubed millimeter. At this
scale the laws of physics are well studied and not related to any global
properties of cosmos, hence, the observed value of cosmological constant does
not match to the known dynamics. It is the essence of cosmological constant
problem reviewed in  \cite{SW} by S.~Weinberg. The conflict with the dynamics
can mean that the observed value of cosmological constant is artificial, i.e.
reducible from a true initial quantity characteristic for the particle
interactions and vacuum structure. In the quantum theory inherently involving
the uncertainty principle, a field cannot be posed in a minimum of its
potential without any motion, so that, by Zel'dovich  \cite{Zel'd}, the
vacuum energy density is related to a non-zero energy for zero number of
quanta of all physical fields, zero-point modes, while a scale of such the
energy should be determined by relevant properties of forces in the nature,
i.e., for instance, by a characteristic scale of grand unification of gauge
interactions about $10^{15}$ GeV  \cite{GUT,PDG}. The extremely small scale
of observed cosmological constant represents itself the problem.

However, since the measured value of vacuum energy density appears as a
reflection of true initial quantity, we have to suppose that this reflection
has been formed during interactions at the same times, when the bare
cosmological constant itself could be essential, i.e. at energy densities of
early Universe. These times refer to a rapid expansion of Universe, which is
called the inflation
 \cite{i-Guth,i-Linde,i-Albrecht+Steinhardt,i-Linde2,inflation} that is
supported by certain real facts in its favor. So, the observed magnitude and
spectrum of anisotropy in the cosmic microwave background radiation
 \cite{WMAP,SPT,Planck,Acbar} can be explained \textit{dynamically}, instead
of introducing an occasional set of very specific initial data of evolution.

In the framework of model for the inflation of early Universe, the anisotropy
is caused by the spatial inhomogeneity of quantum fluctuations of scalar
field called inflaton $\phi$, as calculated in the method of secondary
quantization in vicinity of classical solution for the field
equation\footnote{The dot over a symbol denotes the derivative with respect
to time.} with a potential $V(\phi)$
\begin{equation}\label{field}
    \ddot\phi +3H\dot\phi+\frac{\partial V}{\partial\phi}=0
\end{equation}
in the Universe expanding homogeneously and isotropically with the scale
factor $a(t)$ in the metrics of Friedmann--Robertson-Walker--Lemetre
\begin{equation}\label{FRWL}
    ds^2=dt^2-a^2(t)\,d\boldsymbol r^2,
\end{equation}
with the Hubble parameter $H=\dot a/a$, if the inflaton potential rather
slowly changes near its stable minimum. For instance, in the simplest case of
generic re-normalizable potential
\begin{equation}\label{potenV}
    V(\phi)=\frac{m^2}{8v^2}\,\left(\phi^2-v^2\right)^2,
\end{equation}
with two parameters: mass $m$ and vacuum expectation value $v$, the data
prefer for the scenario of ``new inflation''  \cite{inflation}, when the
field evolves from initial values in the region of potential plateau
$V(0)=m^2v^2/8$ down to the minimum at $\phi=v$ (the ``hilltop'' scenario),
so that we find empirically  \cite{Bstrap}
\begin{eqnarray}
    m\approx (1.52\pm0.22)\cdot 10^{13}\mbox{ GeV}, &&\label{obser-m}
    \\
    2.5\, m_\mathrm{Pl} < v < 54\, m_\mathrm{Pl},\hskip7mm && \label{obser-v}
\end{eqnarray}
where the Planck mass $m_\mathrm{Pl}$ (in units $c=\hbar=1$) is determined by
the Newton gravitational constant $G$ via the relation
$m_\mathrm{Pl}=1/\sqrt{G}=1.22\cdot 10^{19}$ GeV. During the inflation the
Hubble parameter $H$ in the field equation of (\ref{field}) takes the role
analogous to the coefficient of friction in the ordinary mechanics of
point-like body, so that the regime of ``slow roll'' is established for the
field tending to the minimum of potential. In this regime we can neglect the
acceleration of $\ddot\phi$ in (\ref{field}), while $H$ drifts in accordance
to the general relativity with the metrics of (\ref{FRWL}):
\begin{equation}\label{hubble}
      \dot H = -4\pi G\,\dot\phi^2,
\end{equation}
taking into account for the Friedmann equation in the case of matter
saturated by the scalar field, which gives the energy density\footnote{All of
other contributions to the energy density rapidly decline as powers of scale
factor growing during the expansion.} $\rho$,
\begin{equation}\label{Fried}
    H^2=\frac{8\pi G}{3}\rho=\frac{8\pi G}{3}
    \left(\frac12 \dot \phi^2+V(\phi)\right).
\end{equation}

When $H$ is decreasing during the evolution, it crosses a minimal critical
value $H_\mathrm{min}$ depending on the potential model. At lower densities
$H<H_\mathrm{min}$ the field goes out of the slow rolling and starts to
rapidly oscillate in vicinity of potential minimum, with a damping. The
quanta of inflaton vibrations are transformed to quanta of matter field, that
causes the Universe reheating with the substance creation. The matter density
repeats the profile of spatial inhomogeneity of inflaton
 \cite{i-Guth,i-Linde,i-Albrecht+Steinhardt,i-Linde2}.

Like for any observable quantity, we need to quantize the global homogeneous
field of inflaton, which is considered as the classical field in the
description of early Universe, up to the introduction of small spatially
inhomogeneous corrections in the framework of secondary quantization. This
quantization of homogeneous field is necessary in order to clarify the role
of nonzero dispersions of the field and its rate of change onto the run of
Universe evolution. In the present article we consider the averaged quantum
equations of inflaton evolution in the presence of fluctuations. For the sake
of simplicity and clarity of emphasizing physical effects we study the model
with potential
$$
    V=\frac12\, m^2\phi^2.\eqno{(\ref{m2})}
$$
Because of fluctuations, the regime of inflation becomes possible in a
restricted region of field energy density only if the Hubble parameter is
less than a maximal critical value $H_\mathrm{max}$, we find.

\subsection{Quantization and coherent states}

In a comoving volume $V_C$, setting the real physical volume $V_R=a^3V_C$,
the action of homogeneous scalar inflationary field in the metrics of
(\ref{FRWL})
\begin{equation}\label{action}
    S=V_C\int dt \,a^3(t)\left(\frac12 \dot \phi^2+\frac12 m^2\phi^2\right)
\end{equation}
determines the canonical momentum of field
\begin{equation}\label{momentum}
    \hat p_\phi=V_R\dot \phi,
\end{equation}
so that the Hamiltonian
\begin{equation}\label{Hamiltonian}
    \mathcal{H}_\phi=\frac{\hat p_\phi^2}{2V_R}+\frac12 V_R m^2\phi^2
\end{equation}
corresponds to the harmonic oscillator with mass $V_R$ and frequency $m$,
hence, in the oscillatory units of the field $\phi^{(0)}$ and canonically
conjugated momentum  $p_\phi^{(0)}$
\begin{equation}\label{units}
    p_\phi^{(0)}=\sqrt{V_R m},\qquad\phi^{(0)}p_\phi^{(0)}=1,
\end{equation}
the operators of creation and annihilation of field quanta,
$\hat\alpha^\dagger$ and $\hat \alpha$, are defined as
\begin{equation}\label{a}
    \hat\alpha=\frac{1}{\sqrt{2}}\left(\frac{\phi}{\phi^{(0)}}+i\frac{\hat p_\phi}
    {p_\phi^{(0)}}\right).
\end{equation}

In the case of quadratic potential (\ref{m2}), authors of  \cite{Urena} have
offered the description of inflationary dynamics in the method of parametric
attractor in the scaled phase plane of coordinate--momentum, that is
equivalent to the regime of slow rolling in the region of its
applicability\footnote{The generalization of approach of parametric attractor
to the case of potential (\ref{potenV}) has been performed in
 \cite{KT-GREG}.}. In this approach the phase trajectory of inflaton in terms
of scaling variables\footnote{The Friedmann equation of (\ref{Fried}) takes
the form of $x^2+y^2=1$.}
\begin{equation}\label{xy}
    x=\sqrt{\frac{4\pi}{3}}\frac{\dot\phi}{m_\mathrm{Pl}H},\quad
    y=\sqrt{\frac{4\pi}{3}}\frac{m\phi}{m_\mathrm{Pl}H},
\end{equation}
has got stable fixed points, i.e. solutions of equations \mbox{$\dot x=\dot
y=0$,} which positions depend on the Hubble parameter that causes a slow
driftage of field in the phase plane during the inflation. In this way, the
consideration on the phase plane allows us to exactly find the moment of
inflation end due to the constraint on the existence of stable fixed points,
that determines $H_\mathrm{min}=\frac23 m$ as shown in paper \cite{Urena}.

In terms of scaling variables of (\ref{xy}) the operator of quantum
annihilation is reduced to the form of
\begin{equation}\label{ax}
    \hat\alpha=\sqrt{\ell}\,(\hat y+i\hat x),
\end{equation}
where the number of quanta
\begin{equation}\label{nu}
    \ell=\frac{3}{8\pi m}\; H^2m^2_\mathrm{Pl}V_R
\end{equation}
determines the energy spectrum in the volume $V_R$
\begin{equation}\label{E}
    E=m\left(\ell+\frac12\right),
\end{equation}
and the commutator
\begin{equation}\label{comm}
    [\hat y,\hat x]=\frac{i}{2\ell},
\end{equation}
setting the uncertainty relation
\begin{equation}\label{uncert}
    \delta y^2\delta x^2\geqslant \frac{1}{16\ell^2}.
\end{equation}

The energy density of zero-point modes can be easily calculated in terms of
field fluctuations in the vacuum $\delta\dot\phi^2_\mathrm{v}$ and
$\delta\phi^2_\mathrm{v}$,
\begin{equation}
    \langle 0|\hat T_0^0|0\rangle =
    \frac12 \langle 0|(\dot\phi^2+m^2\phi^2)|0\rangle=
    \frac12 (\delta\dot \phi^2_\mathrm{v}+m^2\delta \phi^2_\mathrm{v}),
    \label{vac2}
\end{equation}
or in notations of scaling variables
\begin{equation}
    \langle 0|\hat T_0^0|0\rangle =
    \frac{3}{8\pi}\,H^2m_\mathrm{Pl}^2\,
    \langle 0|(\hat x^2+\hat y^2)|0\rangle=
    \frac{1}{2\ell}\,\rho, 
    \label{vac}
\end{equation}
where we have used the Friedmann equation and elementary identities for the
vacuum
$$
     \langle 0|\hat x^2|0\rangle=\delta x^2_\mathrm{v}=\frac{1}{4\ell},\quad
     \langle 0|\hat y^2|0\rangle=\delta y^2_\mathrm{v}=\frac{1}{4\ell}.
$$
As we should expect, the energy density of zero-point modes in (\ref{vac})
does not depend on time, if we consider the number of quanta in a given
physical volume $V_R$, so that $\ell\sim H^2\sim \rho$ in accordance to
(\ref{nu}).

Note that the energy density of vacuum is composed of contributions by not
only zero-point modes of homogeneous field of inflaton but also by zero-point
modes of secondary quantized inhomogeneous field, too, as well as by
zero-point modes of \textit{all} physical fields. Then, components of
averaged energy-momentum tensor of vacuum fields diverge and one has to
regularize infinities. If for the regularization one uses a method conserving
the isotropy of space-time, for instance, a cut off four-momentum in
four-dimensional Euclidean space, then the contribution of zero-point modes
certainly results in the energy-momentum tensor proportional to the metrics,
as it should be in the case of cosmological constant. Otherwise, if one uses
a regularization conserving only the three-dimensional spatial isotropy, for
instance, a cut off momentum of zero-point modes, then the renormalization of
temporal and spatial components of energy-momentum tensor has to be
considered independently, i.e. it should involve two arbitrary constants. In
the last case, the requirement of space-time isotropy allows us to relate
these two constants of renormalization so that the vacuum tensor of
energy-momentum becomes proportional to the metrics. Therefore, we get the
same result as it has been obtained in the case of isotropic four-dimensional
regularization (see also discussions in  \cite{ACGKS}). In our approach the
finite density of vacuum energy is determined as the energy density
calculated in the framework of canonical quantization of homogeneous inflaton
field:
\begin{equation}\label{vac1}
        \rho_\Lambda^\mathrm{bare}=\langle 0|\hat T_0^0|0\rangle.
\end{equation}

At the inflation start, we suggest that the field and its rate of change
fluctuate eventually, i.e. there is no correlation of its average values
$$
    K=\langle xy+yx\rangle-2\langle x\rangle\langle y\rangle=0.
$$
In quantum mechanics, this situation can be realized for states minimizing
the uncertainty relation for two operators, i.e. for generic coherent states
being eigen-vectors for the operator
$$
    \hat \beta=\frac12\left(\frac{\hat y}{\delta y}+
    i\,\frac{\hat x}{\delta x}\right),
$$
so that $\hat\beta|\beta\rangle=\beta|\beta\rangle$ and $K\equiv 0$ for such
the states. Thus, we conclude that the initial state of Universe before the
inflation has corresponded to the generic coherent state with a maximal
probability.

It is natural to put fluctuations of kinetic and potential energies for the
inflaton to be comparable to each other at the inflation start or
installation, hence, with a high probability $\delta x^2_\mathrm{ins}\sim
\delta y^2_\mathrm{ins}$. Below we will consider quantum equations of
evolution and see that the field evolves slowly, while the evolution of its
rate is very rapid. Therefore, a time, when fluctuations of scaling variables
were exactly equal to each other $\delta x^2=\delta y^2$ in practice
coincides with the inflation start, when the fluctuations were comparable.
Thus, with a high accuracy of leading approximation we can put $\delta
x^2_\mathrm{ins}= \delta y^2_\mathrm{ins}$.

Under the condition of equal fluctuations $\delta x=\delta y$ the operator
$\hat\beta$ transforms into the operator annihilating the oscillator quanta
$\hat \alpha$, while the generic coherent state tends to the coherent state
of oscillator $|\alpha\rangle$.

In the oscillatory coherent state with a complex parameter of average values
$\alpha=\sqrt{\ell}\,(y+ix)$ we get
$ 
    \hat\alpha|\alpha\rangle=\alpha|\alpha\rangle,
$ 
and the fluctuations of scaling variables on the phase plane are determined
by the average amount of quanta
\begin{equation}\label{fluct-xy}
    \delta x^2=\delta y^2=\frac{1}{4\ell},
\end{equation}
exactly the same as in the vacuum. Therefore, the initial state of inflaton
has got the vacuum fluctuations, that agrees with the representation about
the inflation appearance from the state, which contains nothing or almost
nothing except the virtual field and its fluctuations, of course.

The probability versus the number of quanta $k$ is Poisson's distribution
with the average value $\ell$:
\begin{equation}\label{Pois}
    w_k=\mathrm{e}^{-\ell}\,\frac{\ell^k}{k!},
\end{equation}
so that the probability of zero number of field quanta, i.e. the probability
of vacuum, is equal to
\begin{equation}\label{wvac}
    w_\mathrm{vac}=\mathrm e^{-\ell}.
\end{equation}

Note that the energy can be expressed in the form
$$
    E=\rho_\mathrm{ins}V_R+\rho_\Lambda^\mathrm{bare}V_R,
$$
where $\rho_\Lambda^\mathrm{bare}$ is the bare density of zero-point modes,
$\rho_\mathrm{ins}$ is the energy density of inflaton, corresponding to the
installation of inflation in the coherent oscillatory state of inflaton.
Comparing with (\ref{E}), we find
\begin{equation}\label{bare}
    \rho_\Lambda^\mathrm{bare}=\frac{1}{2\ell}\,\rho_\mathrm{ins}.
\end{equation}

\subsection{Evolution of coherent state} Considering the quantized field, we
have to take into account for that the scale factor is the classical
quantity. Therefore, the equation of evolution for the Hubble parameter
should be written in average, i.e.
\begin{equation}\label{hubble2}
      \dot H = -4\pi G\,\langle\dot\phi^2\rangle.
\end{equation}

Introducing $z=m/H$ and the derivative with respect to e-folding of scale
factor \mbox{$N=\ln a/a_\mathrm{ins}$,}
$$
    f^\prime\equiv\frac{df}{dN}=\frac{\dot f}{\dot N}=\frac{\dot f}{H},
$$
where $a_\mathrm{ins}$ is the scale factor at the inflation start, we get the
equation of driftage
\begin{equation}\label{z}
    z^\prime=3\langle{\hat x}^2\rangle\, z.
\end{equation}
For the scaling variables on the phase plane we find
\begin{eqnarray}
  \langle\hat x^2\rangle^\prime &=& -6\langle\hat x^2\rangle+
  6\langle\hat x^2\rangle^2-2z\langle\hat x\rangle\langle\hat y\rangle,
  \label{eq1}\\
  \langle\hat y^2\rangle^\prime &=& 6\langle\hat x^2\rangle
  \langle\hat y^2\rangle+2z\langle\hat x\rangle\langle\hat y\rangle,
  \label{eq2}
\end{eqnarray}
wherein we has taken into account for the coherent states,
$$
    \langle \hat x\hat y+ \hat y\hat
    x\rangle=2\langle\hat x\rangle\langle\hat y\rangle.
$$
It is easy to get $\langle\hat x^2+\hat y^2\rangle^\prime\equiv0$ in
accordance to the Friedmann equation in average.

The inflation corresponds to stable fixed points of system (\ref{eq1}),
(\ref{eq2}), when $\langle\hat x^2\rangle^\prime=\langle\hat
y^2\rangle^\prime=0$, which is possible only if
\begin{equation}\label{zmax}
    z^2\geqslant z^2_\mathrm{min}=36\,\delta x^2(1-\delta x^2)
    \approx\frac{9}{\ell},
\end{equation}
i.e.
\begin{equation}\label{Hmax}
    H^2\leqslant\frac{\ell}{9}\,m^2=H^2_\mathrm{max},
\end{equation}
that can be easily derived by expressing $z$ from the conditions of fixed
points and considering this real parameter as the function of averaged values
of scaling variables.

Thus, the canonical quantization of homogeneous inflaton field allows the
existence of inflation regime in the coherent state, but the quantum
fluctuations restrict the region of inflation development under (\ref{Hmax}).
This is caused by that the average rate of field change $\langle\hat
x\rangle$, corresponding to the fixed point in classics, can be essentially
less than the quantum fluctuations of this rate. At $z=z_\mathrm{min}$ one
get the equality $\langle\hat x\rangle^2=\delta x^2$, while at $z^2\gg
36\,\delta x^2$ the classical limit is reached.

The maximal value of Hubble constant in (\ref{Hmax}) refers to the inflation
installation in the oscillatory coherent state of inflationary field. It
straightforwardly means that
\begin{equation}\label{bare2}
    \rho_\Lambda^\mathrm{bare}=\frac{1}{48\pi}\,
    m^2m^2_\mathrm{Pl}.
\end{equation}
Then, the known estimate of inflaton mass in (\ref{obser-m}) determines the
energy scale of bare density of vacuum energy $\Lambda\approx 4\cdot 10^{15}$
GeV. Such the scale is quite natural for the particle physics in models of
Grand Unification  \cite{GUT,PDG}. In this way, the contribution of bare
cosmological constant in the energy density during the inflation is
suppressed as $1/\ell\ll1$, so that it is inessential at that times.

If we put the maximal accessible value of Hubble parameter to be determined
by the plateau height in the scenario of ``new inflation'', i.e.
$$
    H^2_\mathrm{max}=\frac{\pi}{3m^2_\mathrm{Pl}}m^2v^2,
$$
then
\begin{equation}\label{nuv}
    \ell=3\pi\frac{v^2}{m^2_\mathrm{Pl}}.
\end{equation}


\subsection{Evolution of fluctuations}

At nonzero correlations of scaling variables on the phase plane, equations
for the field fluctuations take the form
\begin{eqnarray*}
  (\delta y^2)^\prime &=\;\;\,& 6\langle\hat x^2\rangle\,\delta y^2
  +z(\langle xy+yx\rangle-2\langle x\rangle\langle y\rangle), \\
  (\delta x^2)^\prime &=-& 6\langle\hat y^2\rangle\,\delta x^2
  -z(\langle xy+yx\rangle-2\langle x\rangle\langle y\rangle).
\end{eqnarray*}
Accounting for the evolution law of Hubble parameter, we deduce that for the
coherent states the fluctuations of field and potential energy slowly grow
with the decrease of Hubble constant as
\begin{equation}\label{dy}
    \delta y^2=\delta y^2_\mathrm{ins}\,\frac{H^2_\mathrm{ins}}{H^2},
\end{equation}
while the fluctuations of field rate and kinetic energy dominantly depend on
the volume $V_R=V_{R,\mathrm{ins}}\, a^3/a^2_\mathrm{ins}$, so that
\begin{equation}\label{dx}
    \delta x^2=\delta x^2_\mathrm{ins}\,\frac{H^2_\mathrm{ins}}{H^2}\,
    \frac{V^2_{R,\mathrm{ins}}}{V_R^2}.
\end{equation}

Since the evolution of kinetic energy fluctuations is exponential with
respect to the amount of e-foldings $N\sim \ln a$, while the evolution of
field fluctuations is the driftage, the moment of fluctuations equality
$N_*$, when $\delta x^2=\delta y^2$, a little bit differs from the moment of
inflation start $N_\mathrm{ins}$, i.e. $|N_\mathrm{ins}-N_*|\ll
N_\mathrm{tot}$. Therefore, the installation of oscillatory coherent state of
inflationary field to the leading approximation practically coincides with
the start of inflation. This fact is enough for our estimates. However, we
have to remember, that further evolution of inflaton state hardly distorts
the wave package in comparison with the moment, when this package was the
oscillatory coherent state.

The total amount of e-folding during the existence of stable fixed point
$$
    N_\mathrm{tot}=\int\limits_{z_\mathrm{min}}^{z_\mathrm{max}}
    \frac{dz}{3\langle\hat x^2\rangle z}
$$
can be easily calculated in the limit $z_\mathrm{min}\ll z_\mathrm{max}$,
since $\langle x^2\rangle=\delta x^2+\langle x\rangle^2$ almost
instantaneously reaches the classical limit $\langle x\rangle^2\approx z^2/9$
because of rapid decline of kinetic energy fluctuations. Thus, for the
potential under consideration in the leading approximation we deduce
\begin{equation}\label{Ntot}
    N_\mathrm{tot}=\frac16\,\ell.
\end{equation}
So, the field fluctuations restrict the total amount of
inflation\footnote{Empirical constraints on $N_\mathrm{tot}$ were studied in
 \cite{LL}.}.

\section{Estimates}

Empirically
\begin{equation}\label{empir}
    \rho_\Lambda=\frac{3}{8\pi}\,H_0^2m_\mathrm{Pl}^2\Omega_\Lambda,
\end{equation}
while in accordance with the data of Planck collaboration  \cite{Planck} the
modern value of Hubble constant is equal to $H_0=67.3\pm 1.2\;
\mbox{km}\,\mbox{sec}^{-1}\mbox{Mpc}^{-1}$, and the fraction of cosmological
constant in the energy balance of flat Universe equals
$\Omega_\Lambda=0.685\,^{+0.018}_{-0.016}$.

On the other hand, the bare value of vacuum energy density in (\ref{bare2})
results in the estimate
\begin{equation}\label{nux}
    \ell=\ln\frac{\rho_\Lambda^\mathrm{bare}}{\rho_\Lambda}=\ln\frac{1}{18}\,
    \frac{m^2}{H_0^2\Omega_\Lambda}\approx 250\gg 1.
\end{equation}
According to (\ref{nuv}) the vacuum expectation value of inflaton equals
\begin{equation}\label{vx}
    v\approx 5.2 \,m_\mathrm{Pl}.
\end{equation}
In this way, we have to note that a further refinement of estimates needs to
make a thorough numerical analysis in more realistic models, say, with
potential (\ref{potenV}), in comparison with the modern data  \cite{Planck}
(see reviews in  \cite{models1,models2}).

Nevertheless, basic physical effects considered in this paper are model
independent: the observed contribution of vacuum energy density is suppressed
by the factor of $\exp\{-\ell\}$ in comparison with the initial bare
cosmological constant, where $\ell$ is the average number of quanta in the
oscillatory coherent state of inflationary field, and the inflation dynamics
is feasible only in the restricted region of energy density because of
quantum fluctuations of the field. An application of particular model allows
us to make comparison with empirical values and to pose constraints of
confidence for a given model of inflation\footnote{In particular, a primary
spectrum of spatial inhomogeneity in the energy density for the simplest
model of quadratic potential (\ref{m2}) disagrees to the measured angular
anisotropy of cosmic microwave background radiation at the level  of
confidence with two standard deviations  \cite{Planck}. In addition, the
total amount of inflationary e-folding $N_\mathrm{tot}$ in (\ref{Ntot}) is
tensely small.}.

\section{Conclusion}

Our result cardinally changes the point of view onto the problem of
cosmological constant
 \cite{SW,Bstrap,Rubakov,SW2,ST,rev-CC,Barr,KV,Chalmers,Grav-seesaw,KT2,ACGKS,KKLT}:
now we understand how the observed artificial scale of vacuum energy density
about $10^{-3}$ eV is deduced from quite the reasonable cosmological constant
with scale of the order of $10^{15}$ GeV because of the mechanism determined
by the installation of initial coherent state for the inflationary field.
Unfortunately, the main parameter of consideration is still empirical, though
it gets the consistent value.

However, we stress that the simplest models of inflationary potential are
marginally in conflict with recent Planck data, while the Starobinsky's model
\cite{Star-1,MCh} is in perfect agreement with those data. Moreover, one has
found the class of models with a nonminimal gravitational interaction of
inflaton, that tends to the fiducial model by Starobinsky in the strong
coupling limit \cite{KLR1}. We expect that the offered mechanism of
cosmological constant installation can be successfully incorporated in that
class of models, while the averaged number of inflaton quanta in the primary
volume could be naturally fixed or predicted. Thus, the account for the
homogeneous quantum fluctuations of inflationary field at the inflation start
makes the inflation model for the early Universe to be more realistic. So,
probably, developing the approach offered in the present paper, we will be
able to prove that this actuality is \textit{dynamically inevitable}.

\acknowledgments
 Author thanks P.~Zheltakova for independent confirmations of
validity for several expressions.

\end{document}